\documentclass{caosp309}

\usepackage{graphicx}

\usepackage{natbib}
\articleNo{B23}
\pubyear{2024}
\volume{54}
\volnumber{1}
\firstpage{1}
\received{November 11, 2023}
\accepted{???}

\begin{document}

\hauthor{H.J.\,Deeg and R.\,Alonso}

\title{Ground-based photometric follow-up for exoplanet detections with the PLATO mission}

\author{
        H.J.\,Deeg\inst{1,2}\orcid{0000-0003-0047-4241}
      \and
        R.\,Alonso\inst{1,2} \orcid{0000-0001-8462-8126}
       }

\institute{
           {Instituto de Astrof\'{\i}sica de Canarias, C. V\'{\i}a L\'{a}ctea S/N, E-38205 La Laguna, Tenerife, Spain}, \email{hdeeg@iac.es}
         \and 
 {Universidad de La Laguna, Dept. de Astrof\'{\i}sica, E-38206 La Laguna, Tenerife, Spain}
          }

\date{November 11, 2023}

\maketitle

\begin{abstract}
Detections of transiting planets from the upcoming PLATO mission are expected to face significant contamination from contaminating eclipsing binaries, resulting in false positives. To counter this, a ground-based programme to acquire time-critical photometry is pursued. Its principal aim is to obtain time-series observations of the planet candidate and its surrounding stars at the times of expected transits. This programme is part of the PLATO Ground-based Observations Programme, which also covers spectroscopic and imaging observations. The current photometric follow-up programme is assembling the required observational resources, executing benchmark observations, and defining strategies for the observations and their reporting. Post-launch, it will focus on coordinating photometric data collection and analysis, and will update candidate statuses in the PLATO follow-up database. Its work packages are outlined, covering specific tools, citizen contributions, standard and multi-colour observations, secondary eclipses, and reprocessing of archival photometry. Ground-based follow-up photometry will likely concentrate on longer-period candidates, given that false positives of short-period candidates will likely become identifiable in timeseries available from GAIA in the near future. Geographical considerations for follow-up observations from the first PLATO long-observation field LOPS2 are outlined, which lies in the southern hemisphere, with later fields expected to be more suitable for northern observers.
\end{abstract}

%
\section{Introduction}
\label{intr}

The succession of space missions dedicated to the detection of transiting exoplanet systems, starting in 2007 with CoRoT, followed by Kepler (and its derivative K2), the current TESS all-sky survey and the forthcoming PLATO mission, has marked an exciting trajectory in the quest for exoplanets. These missions have varied in their approach, with CoRoT and Kepler/K2 conducting deep surveys in relatively small fields, with most transit candidates in the 13th to 15th magnitude range. TESS changed this paradigm with an all-sky survey devoted to stars brighter than 12th magnitudes. The PLATO (PLAnetary Transits and Oscillations of stars) space mission, led by the European Space Agency, primarily aims to discover and study extrasolar planetary systems, with a focus on Earth-like planets in habitable zones around sun-like stars \citep{2014ExA....38..249R,2018haex.bookE..86R}. It seeks to characterize the properties of these exoplanets and their host stars, including their mass, radius, age, and composition. Additionally, PLATO aims to understand stellar evolution and seismic activity by observing stellar oscillations. For these aims, PLATO is set to carry out a deep, wide-field survey focussing on stars brighter than 11th magnitude, with a second large sample of stars up to 13th mag.

For PLATO, a ground-based follow-up observing programme employing spectroscopy, time-critical photometry and imaging forms an integral part of the mission. The role of the ground-based photometric follow-up involves validating the plethora of planet candidates identified in space-based transit surveys, by differentiating genuine exoplanet transits from other astrophysical phenomena. This process usually employs the re-observation of known transit events from ground, but it may also involve multicolour photometric observations or the monitoring of Transit Timing Variations (TTVs) to verify a given candidate (and/or to detect the presence of additional planetary or stellar companions). In this contribution, we describe the current status of the preparation of this follow-up for the PLATO mission.

\begin{figure}
\centerline{\includegraphics[width=1\textwidth,clip=]{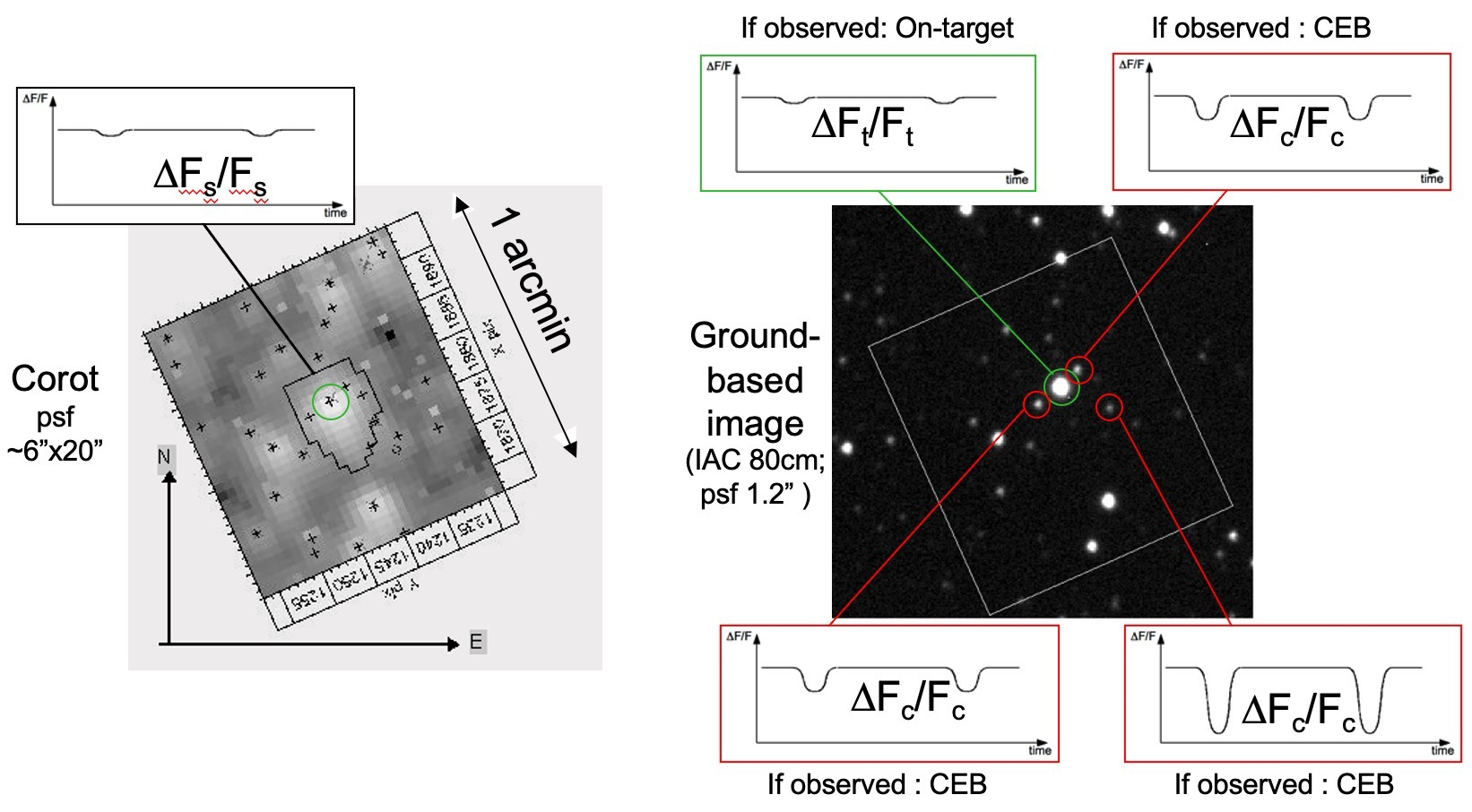}}
\caption{Left: Section of an image from the CoRoT mission around a target star (CoRoT target LRc01\_E1\_2376, marked with a circle). The irregular shape around it indicates the aperture from which photometry is extracted. The crosses indicate catalogued nearby stars. Right: the same field acquired with the IAC 80cm telescope in moderate (1.5") seeing. If the ground observation shows a transit on the target of similar depth (\(\Delta F_t / F_t\), in green box ) as the transit from the space-based light-curve (\(\Delta F_s / F_s\)), the transit is classified as `on-target'. However, if ground-observations show a sufficiently deep eclipse (\(\Delta F_c / F_c\), red boxes) on any of the stars that are close enough to 'contaminate' the target's flux in the aperture, that star is recognised as a CEB. The fact that the detection of a deep eclipse at a nearby star is sufficient to qualify it as a CEB, enables also the follow-up of space-based transits whose depth is too small to be observable as `on-target' transits from ground. Figure adapted from \citet{2009A&A...506..343D}.
}
\label{fig1}
\end{figure}

\section{Identification of false positives from photometric follow-up}
False positives pose a significant challenge in exoplanet transit detections, most often arising from configurations such as grazing and contaminating eclipsing binaries (EBs) within the photometric aperture, which can produce diluted eclipses that superficially resemble planetary transits; see Fig.~\ref{fig1}. The community often employs the terms 'BEB' (blended or background eclipsing binary) or 'CEB' (contaminating eclipsing binary; the term used further in this contribution) to describe these misleading binaries.
The problem of false positives from CEBs was known since the first ground-based transits surveys \citep{2003ApJ...593L.125B}, but got more acute with the launch of the first space mission, CoRoT \citep{2006ESASP1306...33B,2009A&A...506..411A}. That mission's camera had fairly small pixels with an angular side-length of 2.32" when projected into the sky, but its point spread function (psf) was much larger, with a FWHM of about 6" x 20", with the larger value in one direction due to the placement of a prism that permitted the detection of three wavelength-bands. The Kepler mission \citep{2010Sci...327..977B}, launched in 2009, and the `K2' successor mission  \citep[{using the same satellite,}][]{2014PASP..126..398H} had  less issues with CEBs, for one because its image-resolution of 4" reduced the probability of  nearby EBs to fall within the targets' apertures, but also due to the direction of its target field away from the galactic plane, with a lower area density of faint EBs. The TESS mission \citep{2014SPIE.9143E..20R,2015JATIS...1a4003R}, launched in 2018, has four cameras observing a bright sample on adjacent fields, with a relatively low resolution of 21" per pixel. Consequently, it has suffered also a significant contamination by CEBs in its sample of transit detections. The PLATO mission is set to launch in late 2026 and features 24 cameras with  12\,cm apertures (plus 2 `fast' cameras dedicated to bright stars). These cameras are arranged in four groups of six elements, with partially overlapping fields of view, leading to an area of 2232\,deg$^2$ that is covered in each pointing. The PLATO cameras have a pixel size (and approximate resolution) of 15" and will observe fields that include low galactic latitudes, due to which a significant contamination of planet candidates by CEBs is expected. 

The need for a ground-based photometric follow-up effort arose therefore initially with the CoRoT mission, and a dedicated follow-up programme was established for that purpose; its motivation, methodology and early results are described in \citet{2009A&A...506..343D}. A similar effort is currently also under way for the TESS mission, as part of the TESS Follow-Up Observing Program \citep[TFOP,][]{2019AAS...23314005C}. The main observational technique for this follow-up has been the acquisition of single-colour photometric time-series of the planet candidate and nearby stars, at expected transit-times, with the primary aim to validate transit candidates. However, two further photometric techniques are occasionally also of use in the verification and characterisation of transiting planets: Multi-colour photometry, which may indicate false alarms from the typical colour-signatures of  binary eclipses, or it may validate exoplanet transits from their colour signature \citep{2004A&A...425.1125T, 2019A&A...630A..89P}. That validation technique is based on the colour dependence of transits due to stellar limb darkening, which generates specific transit shapes in different wavebands. Multi-colour photometry is mainly useful for the follow-up of deep transits, for which high signal-to-noise observations can be obtained with multi-channel imagers such as the MUSCAT-2 facility of Teide Observatory, Tenerife. Furthermore, observations geared towards the detection of Transit Timing Variations (TTVs) may be useful in refining the orbital parameters and masses of known planets (particularly in multi-planet systems) but also in detecting additional non-transiting planets from their gravitational influence on the transiting ones \citep[e.g.][and references therein]{2018haex.bookE...7A}. 

\section{The PLATO photometric follow-up programme}
The PLATO time critical photometric follow-up programme is part of PLATO's Ground-based Observations Programme (GOP). During the current development /pre-launch phase of PLATO, its principal tasks are: \\
- Estimate the expected demand of observational resources. \\
- Assemble and organize the required observers\\
- Organize and execute benchmark observations to assess the performances of each facility. \\
- Define efficient and effective strategies for time critical follow-up photometry, including standards for the data to be provided and their reporting. \\

In the operational phase, after PLATO's launch, these tasks will change to:\\ 
- Coordinate the collection, analysis and interpretation of time critical photometric data, and the pertaining observing reports.\\ 
- Update the status of observed candidates in the PLATO Follow-Up database after a quality control of the data.\\

This follow-up programme consists of several lower level work packages (WP) under a coordination package (PLATO WP 143000, time-critical photometry from ground), led by R. Alonso at IAC. These lower-level WPs are:
\begin{itemize}
\item WP 143100, Photometric Specific Tools (chair H.J. Deeg, IAC) to define (and develop, if needed) the tools to be used by observers to analyse and report their observations, with reporting standards that are to be defined.  Currently, software packages like AstroImageJ, HOPS, EXOTIC, Siril, MAXIM DL are being evaluated for their utility for the required task, including potential modifications to adapt them to our requirements. These tools might be used by both professional and non-professional contributors, albeit we are aware that professional observers often have their own pipelines in place. These might be used as well, provided that their reporting can be adapted to the required standards.
\item WP 143200, Citizen Contribution (chair G. Wuchterl, Kuffner Sternwarte, Vienna). The PLATO team values the contribution by citizen scientists and amateurs astronomers, and this WP is preparing the interfacing with this community. This includes the development of specific procedures for the dissemination of the targets to be observed and the analysis and reporting of the data. This WP is already performing some test observations of transits  with a group of amateurs, under the label PLATO-Mercury-Test (\url{https://mercurytest.plato-planets.at/}), with a participation open to the interested community.
\item WP143300, Standard and Multi-colour Photometric Observations (chair E. Pall\'e, IAC) is the package which coordinates the observations of time-series of transit candidates by the professional community, both those performing standard observing procedures, and multicolour observations.
\item WP143400, Secondary Eclipses (Chair R. Alonso, IAC) treats special observations for the detection of secondary eclipses (when a planet is occulted by the host star). Of particular interest are cases when these eclipses can be used to recognise false positive scenarios, and/or when the orbital parameters can be improved, as the timing the secondary eclipse provides constraints on the eccentricity.
\item WP143500, Photometry Reprocessing and Homogenisation (chair P. Chote, Warwick Univ., UK) will provide tools and procedures to efficiently access archival time-series photometry acquired by previous surveys, such as TESS, WASP, ASAS, and potentially also photometric time-series from the upcoming GAIA DR4. As a first step, tools are being developed that permit the rapid finding of time-series photometry from a given target within these surveys; the implementation of efficient data-retrieval and eventual reprocessing will be the next development steps.
\end{itemize}

The current preparations of the PLATO follow-up are in defining ground photometry use cases and the procedures and information flows to be implemented, and in organising a collaborative network of observers and observatories. The definition of standards for the data-formats (e.g. meta-data to be provided in FITS headers) and for the reporting of observations is also a current task. A related task will be the development of converters for the ingestion of outputs from various software packages into the PLATO Follow-up Database. A pending issue is also a more detailed evaluation of the impact of photometric time-series from the GAIA DR4, which will be publicly released in late 2025. Using preliminary time-series from GAIA, \citet{2022A&A...667A..14P} were able to identify CEBs within $\approx$5\% of a sample of TESS mission candidates. We therefore expect that GAIA timeseries may provide a significant contribution to the validation of PLATO candidates, with the caveat that PLATO candidates will typically be fainter, but also, that the temporal coverage of GAIA time-series is still increasing, which will improve the detection rates of both on-transit and CEB configurations in GAIA data. It may be expected that most CEBs near 'easy' short-periodic PLATO candidates will become identifiable from GAIA DR4 data, whereas longer-periodic candidates (e.g. periods longer than 20\,d) will continue to need ground-based follow-up.  As a consequence, the second version of the 'Mercury Test', mentioned above for WP 143200, focusses now on candidates with such longer periods. 

For PLATO's first 'Long-duration Observation Phase (LOP), an observing field in the southern hemisphere named LOPS2 has recently been defined, with a field center at a declination of -47.9 $^{\circ}$  (Rauer et al., in prep). Only about 10\% of this field is north of declination -30$^\circ$ and its northernmost edge is a at a declination of about -21$^\circ$. This position implies that LOPS2 cannot be observed meaningfully from anywhere in continental Europe; and only marginally from some other observatories in the northern hemisphere. In the Canary Islands, for example, LOPS2's northernmost tip raises higher than 2 air masses for $\approx$4.5h and reaches a lowest airmass of 1.55. Only from as far south as Hawaii, at a latitude near 20$^{\circ}$N, will it however be possible to observe a substantial fraction ($\approx1/4$) of LOPS2. This means that follow-up for LOPS2 will have to be done nearly exclusively from southern-hemisphere observatories, and efforts to recruit observers for the initial PLATO follow-up will have to concentrate on such locations.
The fields for the remainder of the mission are not defined yet, with a strong contender for the second long pointing being one in the northern hemisphere, named LOPN1, at a declination of +52.9$^{\circ}$ \citep{2022A&A...658A..31N}. Further fields are not defined yet, but they will likely be widely distributed across both hemispheres and consist of shorter pointings, on the order of a few months. Interested observers, independent of their location, are therefore encouraged to sign up for a potential participation in \url{http://tiny.cc/participatePLATOphotFU} or to get in contact with this contribution's authors for further information about the ground-followup for PLATO.

\acknowledgements
We thank the anonymous referee whose comments led to an improvement in the presentation of this work. The authors acknowledge support from the Spanish Research Agency of the Ministry of Science and Innovation (AEI-MICINN) under grant 'Contribution of the IAC to the PLATO Space Mission' with reference PID2019-107061GB-C66, DOI: 10.13039/501100011033.

\bibliography{transit_followup}

\end{document}